\documentclass[showpacs,reprint,amsmath,amssymb,aip,jcp,floatfix]{revtex4-1}

\usepackage{graphicx}% Include figure files
\usepackage{dcolumn}% Align table columns on decimal point
\usepackage{bm}% bold math
\usepackage{amsmath}

\begin{document}

\title{Anomalous molecular orbital variation upon adsorption on wide band gap insulator}
\author{Wei Chen}
\author{Christoph Tegenkamp}
\author{Herbert Pfn\"{u}r}
\email{pfnuer@fkp.uni-hannover.de}
\affiliation{Institut f\"{u}r Festk\"{o}rperphysik, Leibniz Universit\"{a}t Hannover, 30167 Hannover, Germany}
\author{Thomas Bredow}
\affiliation{Institut f\"{u}r Physikalische und Theoretische Chemie, Universit\"{a}t Bonn, 53115 Bonn, Germany}

\date{\today}

\begin{abstract}
It is commonly believed that organic molecules are physisorbed on the ideal non-polar surfaces of wide band gap insulators with 
limited variation of the electronic properties of the adsorbate molecule.
On the basis of first principles calculations within density functional theory (DFT) and $GW$ approximation, we show that this is not generally true.
We find that the molecular frontier orbitals undergo significant changes when a hydroxy acid (here we chose gluconic acid) is adsorbed on MgSO$_4$$\cdot$H$_2$O(100) surface due to the complex interaction between the molecule and the insulating surface.
The predicted trend of the adsorption effect on the energy gap obtained by DFT is reversed when the surface polarization effect is taken into account via the many-body corrections.

\end{abstract}

\pacs{71.15.Mb, 68.43.Bc, 73.20.Hb}
\maketitle

%\begin{tocentry}
%\includegraphics{TOC}
%adsorption, insulator, rehybridization
%\end{tocentry}

Organic-insulator systems are regaining increasing interest in recent years stimulated by the rapid development of novel molecular-scale devices \cite{Dienel2008}.
The performance of these molecular electronics relies intimately on the properties of the frontier orbitals of the organic molecule \cite{Song2009}
and the coupling of the orbitals to the surface electronic states. 
From this point of view, an accurate control of the electrical characteristics requires unambiguous knowledge of the interaction between the molecular adsorbate and insulating surface.

The perfect non-polar surface of a wide band gap insulator, e.g. alkali halides is chemically inert with respect to organic molecules,
and the features of the molecular orbitals are usually preserved upon adsorption as resolved by scanning tunneling spectroscopy \cite{Repp2005} and first principles calculations \cite{Chen2010}.
The ability to decouple adsorbed molecules electrically from the insulating surface also facilitates the use of alkali halides as supporting templates for a wide range of applications \cite{Tegenkamp2002,Mativetsky2007,Pakarinen2009}.
However, as we will show in this study, this scenario does $not$ necessarily hold for all organic-insulator systems.
In contrast, we find that, using density functional theory (DFT), the frontier orbitals of an adsorbate can vary significantly upon adsorption on the surface of a wide band gap insulator.
Here we choose gluconic acid (GA) as a representative for the chemical class of hydroxy acids adsorbed on MgSO$_4$$\cdot$H$_2$O(100) as the subject of investigation.
Apart from the general relevance of the adsorption properties of this wide spread class of molecules, our study is motivated also by the ability of this molecule to act as a conditioner for the electrostatic separation of mineral \cite{Tegenkamp2002}. 
We show that the strong variations of the frontier orbitals arise from the complex interactions between the adsorbate and the underlying surface.
Further, the polarization effect, which is induced by changes in the charge state of the molecule, produces a prominent reduction of the molecular energy gap upon adsorption.
The description of the polarization invokes the use of $GW$ approximation \cite{Hedin1965} since this nonlocal correlation effect is not captured by Kohn-Sham (KS) DFT eigenvalues with standard exchange-correlation (XC) approximations.
We stress that the results, although based on the GA-MgSO$_4$$\cdot$H$_2$O system, provide a general insight into the interaction of a wide class of organic molecules (e.g. carboxylic and phenylic acids) on insulating surfaces.

We start with a brief description of the bulk MgSO$_4$$\cdot$H$_2$O.
MgSO$_4$$\cdot$H$_2$O has a monoclinic structure (space group $C_{2h}^6$), and is chemically formed from MgSO$_4$ by incorporating one water molecule per unit.
The crystal structure parameters were calculated with the PBE XC potential \cite{Perdew1996} within the generalized gradient approximation (GGA) as implemented in \textsc{vasp} \cite{Kresse1996}. 
The electron-ion interaction was described within the projector augmented wave method \cite{Kresse1999}.
A kinetic energy cutoff of 500 eV and a $5\times5\times5$ $\mathbf{k}$-point mesh in the Brillouin zone were used.
The convergence criterion of lattice parameter and atomic position relaxation was set to 0.02 eV/\AA.
The resultant lattice constants ($a=6.79$ \AA, $b=7.79$ \AA, $c=7.69$ \AA) and $\beta=117.7^\circ$ are in good agreement with experiment and previous theoretical values \cite{Maslyuk2006}.
The direct band gap at the $\Gamma$ point is severely underestimated by the GGA-PBE (5.53 eV) and the local density approximation (LDA) (5.26 eV) compared to a surface sensitive experimental value (7.4 eV) \cite{Maslyuk2005}.
A much more realistic gap of 7.41 eV is obtained with the screened hybrid functional HSE06 \cite{Heyd2003,Krukau2006} owing to the alleviation of the self-interaction error and derivative discontinuity problem by the nonlocal exact exchange.
As resolved from the projected density of states (PDOS) (not shown), the valence band maximum (VBM) is of O-$2p$ character and the conduction band minimum consists of Mg-3$s$ and S-$3s$ states.
In particular, a pronounced broadening can be seen in the water $1b_1$ orbital as a consequence of the bonding to the neighboring Mg atoms.

\begin{figure}
\includegraphics{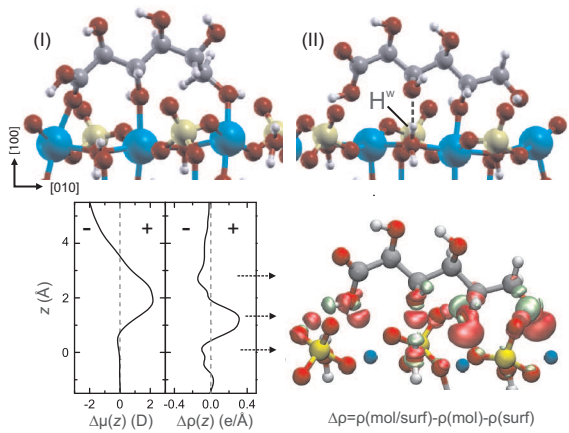}
\caption{\label{Ads_Conf}Top: Two representative configurations of GA adsorbed on MgSO$_4$$\cdot$H$_2$O(100).
Color code: Mg blue (large gray), S yellow (light gray), O red (black), C small gray, H white.
Bottom: Adsorbate induced electron density difference $\Delta\rho$ for Conf.~II obtained from HSE06 calculations along with the planar integrated $\Delta\rho(z)$ and accumulated induced dipole $\Delta\mu(z)$. 
The red (dark gray) and green (light gray) regions correspond to electron accumulation and depletion, respectively. The isosurface value is $\pm$0.03 $e$/\AA$^3$.}
\end{figure}

\begin{table}
\caption{\label{energy} Adsorption ($E_\text{ads}$) and interaction energy ($E_\text{int}$) (in eV) for the two adsorption configurations optimized by PBE-D as shown in Fig.~\ref{Ads_Conf}.
The GGA-PBE contributions to the $E_\text{ads}$ ($E_\text{int}$) are explicitly listed. The deformation energy of the molecule is denoted by $\Delta$$E_\text{d}$.}
\begin{ruledtabular}
\begin{tabular}{lccccc}
Conf. & $E_\text{ads}^\text{PBE-D}$ & $E_\text{ads}^\text{PBE}$ & $E_\text{int}^\text{PBE-D}$ & $E_\text{int}^\text{PBE}$ & $\Delta$$E_\text{d}$ \\
\hline
I & -2.47 & -1.63 & -4.26 & -3.42 & 0.77\\
II& -2.49 & -1.53 & -4.04 & -3.09 & 0.53\\
\end{tabular}
\end{ruledtabular}
\end{table}

For the adsorption of GA, a (1$\times$2) MgSO$_4$$\cdot$H$_2$O(100) surface supercell with 14 atomic layers (72 atoms) was adopted. 
Test calculations using larger unit cells confirmed that the (1$\times$2) supercell is adequate for various adsorption configurations while keeping the intermolecular interaction minimized.
By construction, the (100) surface is cleaved in a way that the electrostatic dipoles between the Mg$^{2+}$ and SO$_4^{2-}$ are nearly within the surface plane.
Hence, the macroscopic dipole moment is negligible along the surface normal, making this surface thermodynamically more favorable than other surface planes.
The non-polar nature of the surface is also evidenced by the small displacements of the surface atoms during relaxations.
For detailed description of the (100) surface the reader is referred to Ref.~\citenum{Maslyuk2005}.
A vacuum thickness of 14 \AA\ was used to separate the adsorbate system from its periodic image.
The calculations have been checked to converge well with respect to the slab thickness, $\mathbf{k}$-point mesh ($2\times4\times1$) and kinetic energy cutoff (400 eV).
For geometrical relaxations, we employed the PBE-D scheme \cite{Grimme2006,Chen2010} to take into account the long-range van der Waals (vdW) force where it is important \cite{Chakarova-Kack2006,Chen2009,Pakarinen2009}. 
In PBE-D, the dispersion (D) is constructed by an empirical $C_6R^{-6}$ potential, which allows for a transparent and fast assessment of the vdW interactions in standard DFT approximations.
The electronic properties of the adsorbate system were then computed with the HSE06 hybrid functional.

It is evident from the rotational flexibility of the GA molecule that the geometry optimizations of the adsorbate system will lead to numerous stable or meta-stable configurations and it is computationally prohibitive to locate all the local minima.
In the present study, we used a scanning scheme in which the GA molecule was initially placed at several different positions on the (1$\times$2) surface supercell with an orientation parallel to the surface plane.
The whole adsorbate system was then allowed to fully relax until the equilibrium structure is reached, and the adsorption and interaction energy can be calculated as
\begin{equation}
E_\text{ads}=E_\text{system}^\text{relaxed}-E_\text{surface}^\text{relaxed}-E_\text{molecule}^\text{relaxed}
\end{equation}
and
\begin{equation}
E_\text{int}=E_\text{system}^\text{relaxed}-E_\text{surface}^\text{ads}-E_\text{molecule}^\text{ads},
\end{equation}
where $E^\text{relaxed}$ denotes the energy of the corresponding unit in its equilibrium geometry, and $E^\text{ads}$ is the energy of the unit in the adsorbate geometry.
The difference between $E_\text{ads}$ and $E_\text{int}$ is the deformation energy of the molecule and surface upon interaction.
Dissociative adsorption via deprotonation has been found energetically unfavorable and is not included in this study.
We find that the PBE-D method yields a nearly identical adsorption geometry as the PBE except that the dispersion force pulls the molecule 0.1 \AA\ closer to the surface.
This suggests that the adsorption configuration is largely driven by the short-range interactions.
In Fig.~\ref{Ads_Conf}, the two configurations with the lowest $E_\text{ads}$ (see Table~\ref{energy}) are presented, both of which depict the fixture of the molecule via multiple localized Mg-O and O-H bonds.
Such bonding type implies that an ordered structure can be achieved on MgSO$_4$$\cdot$H$_2$O(100) where the GA molecule is prone to lie flat on the surface along the [010] direction.
Specifically, in Conf.~I the molecule is stabilized via three Mg-O bonds from the carboxylic and hydroxylic O atoms with bond lengths ranging from 2.10 to 2.16 \AA.
The incorporated water molecule at the surface is not interacting with the adsorbate.
In the second configuration, we see a dominating Mg-O bond from the hydroxylic O (2.06 \AA), accompanied by two weaker Mg-O bonds with larger bond lengths (2.22 and 2.55 \AA).
Besides, one of the hydroxylic O atom binds to the hydrogen of the water molecule at the surface (H$^\text{w}$).
The difference in adsorption configurations has a direct influence on the binding energy as shown in Table~\ref{energy}, where $E_\text{int}$ calculated with GGA-PBE is 0.33 eV higher for Conf.~I.
While the vdW interaction ($E_\text{int}^\text{PBE-D}-E_\text{int}^\text{PBE}$) has been found substantial for both configurations (0.84 and 0.95 eV),
it accounts for a relatively small portion (20-25\%) of the total $E_\text{int}$ compared to the short-range molecule-surface interactions.
Hence, we conclude that the long-range vdW force does not play a predominant role in the adsorption of GA on MgSO$_4$$\cdot$H$_2$O(100).

\begin{figure}
\includegraphics{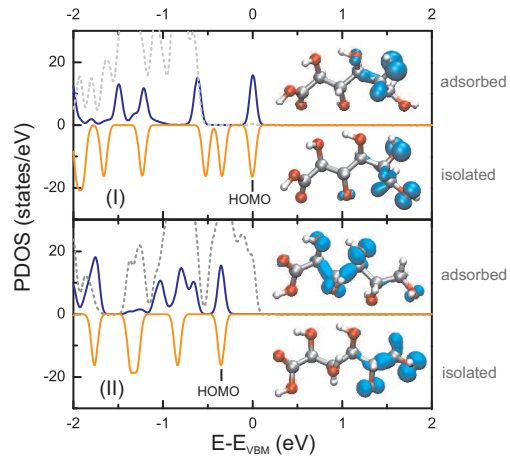}
\caption{\label{DOS} Density of states projected onto the GA molecule (adsorbed and isolated) (solid line) and surface (gray dashed line) calculated with the HSE06 functional for Conf.~I and II.
A Gaussian smearing of 0.05 eV has been applied. The HOMO of the isolated molecule is aligned to that of the adsorbate.
The electron density distributions of HOMO for the adsorbed and isolated molecules are also illustrated.}
\end{figure}

Inspecting the charge density difference (Fig.~\ref{Ads_Conf}), one finds electron accumulations in between the bond region upon adsorption, as expected for covalent interactions.
Note that the electron redistribution at the molecule-surface interface induces a change in the surface dipole by -2.3 D. 
The covalent character is also seen in Fig.~\ref{DOS}, where the molecular orbitals of the adsorbate exhibit broadening and splitting due to rehybridizations.
This can be rationalized as a dative covalent bond through the donation from the lone-pair electrons of the O atoms.
However, the covalent bonds do not contribute to the large interaction energy from the GGA-PBE part.
In contrast, they act as a repulsive Pauli barrier because both the bonding and antibonding states are fully occupied.
Therefore, we ascribe the strong interaction energy to the attractive electrostatic interaction arising from the Coulomb terms.
The covalent character, on the other hand, accounts for the electronic structure of the adsorbate as we will show in a moment.
Furthermore, the charge transfer is found to be small.
For instance, a Bader analysis \cite{Tang2009} predicts that 0.01 $e$ are transferred to the molecule for Conf.~II.
This is expected as the energy gaps of both the surface and molecule are rather large.

An interesting indication of the large $\Delta$$E_\text{d}$ in Table \ref{energy} is the pronounced intramolecular distortion from its equilibrium geometry upon adsorption.
This feature stems partially from the intrinsic molecular structure of GA, and is not observed for some other organic molecules, e.g. hydroxybenzoic acid on MgSO$_4$$\cdot$H$_2$O(100) \cite{Maslyuk2006}.
The $sp^3$ hybridization in the carbon chain makes the structure versatile through the rotation along the C-C bond with a small energy barrier, which can be easily overcome by the energy gain through the subsequent GA-surface interactions. An immediate consequence of this structural change can be manifested by the reduced gap of the adsorbate with respect to that of the gas phase molecule (Table~\ref{gap}).
In addition, our calculation on KCl(001) shows a minor deformation of the GA molecule with a much weaker $E_\text{ads}$ (-0.4 eV excluding the vdW force), an indication that the intramolecular distortion is also related to the surface.
The higher reactivity of MgSO$_4$$\cdot$H$_2$O(100) is associated with its lower Madelung potential as the ions at the surface are more exposed to the environment.

\begin{table}
\caption{\label{gap} Calculated energy gaps (in eV) of the GA molecule by DFT and $GW$ approximation.
The adsorbed molecule refers to the GA molecule adsorbed on MgSO$_4$$\cdot$H$_2$O(100) as in Conf.~II, whereas the isolated molecule refers to the GA molecule detached from the surface while its geometry is kept fixed as that of the adsorbed molecule.
The GA molecule in its equilibrium geometry is denoted by the gas phase molecule.
The energy gaps of the SA adsorbed on MgSO$_4$$\cdot$H$_2$O(100) are given in parentheses for comparison.}
\begin{ruledtabular}
\begin{tabular}{lcccc}
 &PBE & HSE06 & LDA & \multicolumn{1}{c}{$G_0W_0^\text{LDA}$} \\
\hline
adsorbed & 4.84 (3.21) & 6.73 (4.30) & 4.81 & 10.01\\
isolated & 4.10 (3.37) & 5.97 (4.49) & 4.06 & 10.59\\
gas phase& 4.46 (3.17) & 6.39 (4.22) & 4.43 & 10.94\\
\end{tabular}
\end{ruledtabular}
\end{table}

\begin{figure}
\includegraphics{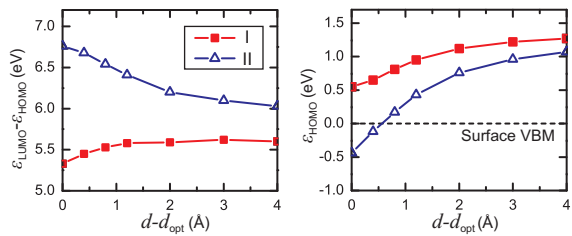}
\caption{\label{MO} The evolution of the HOMO-LUMO gap (left) and the HOMO energy (right) (within HSE06) of the GA adsorbate with respect to the distance to the surface $d$. $d_\text{opt}$ is the separation between the molecule and surface at the optimized geometry.}
\end{figure}

We now turn to the surface and adsorbate induced effect on the molecular orbital of the GA molecule. 
It is surprising to see from Fig.~\ref{DOS} that, in Conf.~II, the highest occupied molecular orbital (HOMO) experiences dramatic changes upon adsorption, even with the intramolecular structure kept intact.
A detailed analysis reveals that the order of the HOMO and HOMO-1 is reversed after the molecule is attached to the surface.
Consider the HOMO and HOMO-1 of GA in the gas phase are by no means degenerate, such \textit{reordering} of the molecular orbitals is rather \textit{unusual} on wide band gap insulators.
This sends a signal that the adsorption can strongly modify the molecular orbitals even on pristine insulating surfaces.
The interchange of HOMO and HOMO-1 subsequently increases the energy gap of the adsorbate by about 0.75 eV with respect to the isolated molecule (see Table~\ref{gap}). 
We note that such gap variation shows no dependence of the XC functional used.
The drastic modification of the HOMO, however, is absent for Conf.~I where the electron density redistribution is considerably smaller.
In Fig.~\ref{MO} two distinct behaviors of the frontier orbitals are perceivable when the GA molecule is lifted away from the surface while the intramolecular structure is fixed at its adsorbate state.
In Conf.~I, the gap between the HOMO and the lowest unoccupied molecular orbital (LUMO) experiences a small decrease relative to the isolated molecule.
As the molecular HOMO in Conf.~I lies within the surface band gap, there is no coupling between the HOMO and the surface valence band.
Hence, the small gap reduction arises from the electrostatic potential of surface dipoles.
This also holds true for the hydroxybenzoic acid (SA) adsorbate on MgSO$_4$$\cdot$H$_2$O(100) as well as the GA on alkali halide (001) surface.
On the other hand, for Conf.~II, the HOMO of the adsorbate is pinned below the surface VBM as a result of the resonance between the hydroxylic oxygen and H$^\text{w}$ of the surface.
Since the hydrogen bond weakens rapidly as the molecule is gradually detached from the surface, one can see the accelerated upshift of the HOMO against the VBM as well as the declining of the molecular energy gap in Fig.~\ref{MO}. 
It is thus clear that the strong variations of the frontier orbitals upon molecular adsorption stem from the rehybridizations.
The water molecule, which was usually thought to be inactive in MgSO$_4$$\cdot$H$_2$O, however plays an important role in the electronic properties of the adsorbate.

We have seen up to this point that the frontier orbitals of the adsorbate can be effectively influenced by the molecule-surface interactions based on the interpretation from the mean field KS equation within the single particle Hamiltonian.
Recently it has been found that dynamic polarization effect has a strong impact on the energy level position of the frontier orbital by photoemission \cite{Johnson1987} and electron transport measurements \cite{Kubatkin2003,Repp2005}. 
The polarization in the surface, which is induced by the added electron (hole) in the affinity (ionization) level, acts back to the molecule and gives rise to a renormalization of the molecular levels.
Apparently such nonlocal correlation is beyond the scope of KS-DFT and is not properly represented by KS eigenvalues, while it is accessible from the self-energy $\Sigma=iGW$ within many-body corrections, where $G$ and $W$ are the Green's function and the screened Coulomb interaction, respectively \cite{Neaton2006,Garcia-Lastra2009,Freysoldt2009}.
To illustrate to what extent the affinity and ionization levels of the GA adsorbate are shifted due to the polarization, $G_0W_0$ calculations \cite{Hedin1965,Aryasetiawan1998} on top of LDA are performed for Conf.~II.
We note that the polarization effect should be similar for Conf.~I.
The $G_0W_0$ approach is a first-order perturbation to the KS eigenvalues, and the quasiparticle (QP) correction to the energy level can be written as the difference between the self-energy and the XC potential $V_\text{XC}$.
We used \textsc{quantum-espresso} \cite{Giannozzi2009} for the LDA, followed by $G_0W_0$ calculations with \textsc{yambo} \cite{Marini2009}. Norm-conserving pseudopotentials, a cutoff energy of 816 eV (60 Ry) and a $2\times4\times1$ $\mathbf{k}$-point mesh were used. We included 200 and 400 empty bands for the isolated GA and adsorbate system in the evaluation of self-energy $\Sigma$, respectively.\footnote{The choice of empty band is justified as the convergence test showed that the QP energy gaps changed by 0.02 eV when 400 and 800 empty bands were used for the GA molecule and adsorbate system, respectively.} 
For the dielectric matrix 3400 reciprocal lattice vectors were used. 
The random phase approximation (RPA) and the plasmon-pole approximation with a plasmon frequency of 27.2 eV were applied for the response function.
The slowly decaying Coulomb potential in the repeated-slab approach is corrected with a boxlike cutoff.
The QP gap was converged within 0.1 eV with these parameters.

The resulting QP gap from the $G_0W_0^\text{LDA}$ for the gas phase GA exhibits a much larger opening than the HSE06 HOMO-LUMO gap (see Table~\ref{gap}). 
When the molecule is brought into contact with the surface, a pronounced gap reduction (0.58 eV) is obtained from the $G_0W_0^\text{LDA}$ calculation.
This is clearly opposed to what has been found by DFT calculations.
As the charge transfer is very small, the gap reduction upon adsorption from the $GW$ approximation arises from the surface polarization effect, which can be described by the classical image charge theory \cite{Rohlfing2003,Neaton2006,Garcia-Lastra2009}.
The affinity (ionization) level of the molecular adsorbate moves down (up) by the image charge potential 
%\begin{equation}
%\label{Vim}
$V_\text{im}=\frac{1}{4}\frac{\epsilon-1}{\epsilon+1}\frac{e^2}{z-z_0}$,
%\end{equation}
where $\epsilon$ is the dielectric constant of the medium, $z$ and $z_0$ the position of the point charge and image plane, respectively.
This implies that relative to the isolated molecule, the changes in the QP correction to the HOMO and LUMO should be antisymmetric.
In fact, we find that $\Delta$$E_\text{HOMO}^\text{QP}=0.6$ eV and $\Delta$$E_\text{LUMO}^\text{QP}=-0.7$ eV, corresponding to a gap reduction of 1.3 eV due to surface polarization.
Note that apart from the polarization effect, the change in the energy gap upon adsorption given by $G_0W_0$ ($\Delta E_g^{G_0W_0}$) in Table~\ref{gap} also includes the contribution from the local interactions 
\begin{equation}
\Delta E_g^{G_0W_0} = \Delta E_g^\text{pol} + \Delta E_g^\text{local},
\end{equation}
where $\Delta E_g^\text{pol}$ is the polarization-induced change in the energy gap, and $\Delta E_g^\text{local}$ refers to the energy gap variation due to short-ranged effects such as electrostatic interactions and rehybridizations, which should be in principle accurately described at the DFT level.
Indeed, we find that $\Delta E_g^\text{local} = (-0.58)-(-1.3)=0.72$ eV, in excellent agreement with the HOMO-LUMO gap change upon adsorption from the DFT calculations (see Table~\ref{gap}).  

In summary, we have demonstrated that upon adsorption to the surface of a wide band gap insulator, the molecular orbitals of the adsorbate can experience substantial changes as a result of the complex interplay of the sizeable electrostatic interaction and rehybridization. 
Further, the polarization effect as described by the $GW$ approximation, yields a prominent molecular energy gap reduction on the wide band gap insulator.
As the electronic properties are dictated by the energetic positions and characteristics of frontier orbitals of the adsorbate, one has to carefully assess the molecule-insulator interface when interpreting the transport and spectroscopic measurements.

\begin{acknowledgements}
We thank H\"ochstleistungsrechenzentrum Nord (HLRN) for the generous grants of computation time. This work is supported by K+S AG.
\end{acknowledgements}

%\bibliography{GA_MgSO4H2O}

%

\end{document}